\documentclass[]{raa}            
\usepackage{graphicx,times}
\usepackage{natbib}

\providecommand{\bm}[1]{\mbox{\boldmath$#1$\unboldmath}}

\begin{document}

   \title{Binary population synthesis study on supersoft X-ray phase of single
degenerate type Ia supernova progenitors
}

 \volnopage{ {\bf 2009} Vol.\ {\bf 9} No. {\bf XX}, 000--000}
   \setcounter{page}{1}

   \author{Xiang-Cun Meng
      \inst{1}
   \and Wu-Ming Yang
      \inst{1,2}
   }

   \institute{School of Physics and Chemistry, Henan Polytechnic
University, Jiaozuo, 454000, China; {\it xiangcunmeng@hotmail.com}\\
        \and
             Department of Astronomy, Beijing Normal University, Beijing 100875, China
}

\abstract{In the single degenerate (SD) scenario for type Ia
supernovae (SNe Ia), a mass-accreting white dwarf is expected to
experience a supersoft X-ray source (SSS) phase. However, some
recent observations showed that the expected number of the
mass-accreting WD is much lower than that predicted from theory,
whatever in spiral or elliptical galaxies. In this paper, we did a
binary population synthesis study on the relative duration of the
SSS phase to their whole mass-increasing phase of WDs leading to
SNe Ia. We found that for about 40\% progenitor systems, the
relative duration is shorter than 2\% and the evolution of the
mean relative duration shows that it is always smaller than 5\%,
whatever for young or old SNe Ia. In addition, before SNe Ia
explosion, more than 55\% progenitor systems are experiencing a
dwarf novae phase, and only no more than 10\% is staying SSS
phase. These results are consistent with the recent observations,
and imply that both in early- and late-type galaxies, only a small
fraction of mass-accreting WD resulting SNe Ia contribute to the
supersoft X-ray flux. So, although our results are not directly
related to the X-ray output of SN Ia progenitor, the low supersoft
X-ray luminosity observed in early type galaxies may have no
ability to exclude the validity of SD model. On the contrary, it
is evidence to support the SD scenario.
  }

   \authorrunning{Meng \&  Yang}            
   \titlerunning{Binary population study on supersoft X-ray}  
   \maketitle


%
%
\section{INTRODUCTION}           
\label{sect:1} Although Type Ia supernovae (SNe Ia) are very
important in astrophysics, i.e. as standard candles to measure
cosmological parameters(\citealt{RIE98}; \citealt{PER99}), the
exact nature of their progenitors is still unclear
(\citealt{HN00}; \citealt{LEI00}; \citealt{PAR07}). There is a
consensus that SNe Ia result from the thermonuclear explosion of a
carbon--oxygen white dwarf (CO WD) in a binary system
(\citealt{HF60}). According to the nature of the companions of the
mass accreting WDs, two basic scenarios for the progenitors of SN
Ia have been discussed over the last three decades. One is the
single degenerate (SD) model (\citealt{WI73}; \citealt{NTY84}),
i.e. the companion is a main-sequence or a slightly evolved star
(WD+MS), a red giant star (WD+RG) or a helium  star (WD + He star)
(\citealt{LI97}; \citealt{HAC99a}; \citealt{LAN00};
\citealt{HAN04}; \citealt{CHENWC07}; \citealt{MENG09};
\citealt{LGL09}; \citealt{WANGB09a}; \citealt{MENGYANG10c};
\citealt{WANGB10}). The other is the double degenerate (DD) model,
i.e. the companion is another CO WD (\citealt{IT84};
\citealt{WEB84}).

It is believed that the X-ray signature of these two possible
scenarios is very different, i.e. no strong X-ray is expected from
the DD scenario until shortly before SNe Ia explosion, while the
mass-accreting WDs may show the signatures of supersoft X-ray
sources (SSSs) long before it explodes as SNe Ia. This provides a
potential method to distinguish the SD and DD model. Recently,
\citet{GB10} (hereafter GB10) obtained the 0.3-0.7 keV soft X-ray
luminosity and the K-band luminosity of several early type
galaxies. By comparing the detected X-ray luminosity with those
predicted from the SNe Ia rate estimated from the K-band
luminosity, they found that the observed X-ray flux from the early
type galaxies is a factor of $\sim$ 30 - 50 less than predicted
from the SD accretion scenario, and then they concluded that no
more than five percent of SNe Ia in early type galaxies can be
produced by mass-accreting white dwarfs of the SD scenario.
However, as noticed by \citet{HKN10}, the results of GB10 based on
an assumption that all the accreting WDs are in the SSS phase
lasting for about two million years before SNe Ia explosion. This
assumption is rather simple and may overestimate the duration of
SSS phase by a factor of 10. \citet{HKN10} recalculated their
symbiotic model and found that the results in GB10 strongly
supports the SD scenario as a main contributor of SNe Ia in early
type galaxies. Similar to GB10, \citet{DISTEFANO10} also noticed
that the predicted SSS number from SD model in spiral galaxies is
also much larger than that from observations, which may imply that
the duration of SSS phase from SD scenario is much shorter than
the accreting time of WD, as presented in \citet{HKN10}. Actually,
many works have referred to this problem, i.e the duration of SSS
phase for an accreting WD is very short (see \citealt{YUN96};
\citealt{HAN04}; \citealt{MENG09}; \citealt{MENGYANG10a}). In
\citet{MENGYANG10a}, they calculated a dense model grid including
WD + MS and WD + RG systems, and the calculations obtained a very
short duration of SSS phase for an accreting WD, i.e. the typical
value is $1.5-4.5\times10^{\rm 5}$ years which is consistent with
the calculation of \citet{HKN10} (see figures 1 and 2 in
\citealt{MENGYANG10a}). In this paper, we want to use the
calculations in \citet{MENGYANG10a} to do a binary population
synthesis study to check whether or not the SD model in
\citet{MENGYANG10a} may explain the discoveries of
\citet{DISTEFANO10} and GB10.

In section \ref{sect:2}, we simply describe our method, and
present the calculation results in section \ref{sect:3}. In
section \ref{sect:4}, we show discussions and our main
conclusions.


\section{METHOD}\label{sect:2}
Recently, \citet{MENGYANG10a} constructed a comprehensive single
degenerate progenitor model for SNe Ia. In the model, the
mass-stripping effect of optically thick wind (\citealt{HAC96})
and the effect of a thermally instable disk were included
(\citealt{HKN08}; \citealt{XL09}). The prescription of
\citet{HAC99a} on WDs accreting hydrogen-rich material from their
companions was applied to calculate the WD mass growth. The
optically thick wind and the material stripped-off by the wind
were assumed to take away the specific angular momentum of the WD
and its companion, respectively. In \citet{MENGYANG10a}, both WD +
MS channel and WD + RG channel are considered, i.e. Roche lobe
overflow (RLOF) begins at MS or RG stage. The Galactic birth rate
of SNe Ia derived from that model is comparable with that from
observations. In addition, that model may even explain some
supernovae with low hydrogen mass in their explosion ejecta
(\citealt{MENGYANG10b}). In their calculations, the progenitor
system may experience a wind (Case wind), a calm hydrogen-burning
(Case Calm), a nova explosion (Case Nova) or a dwarf nova phase
(Case DNova). If a system is undergoing a calm hydrogen-burning
phase, it may show the properties of SSSs. \citet{MENGYANG10a}
calculated more than 1600 WD close binary evolutions and showed
the initial parameter space leading to SNe Ia in an orbital period
- secondary mass ($\log P_{\rm i}, M_{\rm 2}^{\rm i}$) plane.
Their calculation is comprehensive, and the evolution details of
all the models are obtained but not sorted for publishing. In this
paper, we extract the evolutionary properties of the models from
the data files of the calculations and incorporate them into the
BPS code developed by \citet{HUR00, HUR02} to obtain the relative
duration of SSS phase to the whole mass-increasing time of WD
leading to SNe Ia.

We simply describe the assumptions for mass accretion in
\citet{MENGYANG10a}. In an initial binary system, the companion
fills its Roche lobe at MS or during HG or RG, and mass transfer
occurs. They assume that the transferred materials form a disk
around the CO WD. If the mass-transfer rate is lower than a
critical mass-transfer rate, the disk is thermal unstable and a
dwarf nova is expected, otherwise the WD accretes the transferred
material smoothly at a rate $\dot{M}_{\rm a}=|\dot{M}_{\rm tr}|$,
where the critical mass-transfer rate is a function of the orbital
period of binary system:
 \begin{equation}
 \dot{M}_{\rm c, th}=4.3\times 10^{\rm -9}(\frac{P_{\rm orb}}{4{\rm hr}})^{1.7} M_{\odot}{\rm yr}^{\rm -1},
  \end{equation}
where $P_{\rm orb}$ is the orbital period unit in hour
(\citealt{OSAKI96}; \citealt{VANP96}). If the accretion rate of a
CO WD, $\dot{M}_{\rm a}$, is larger than a critical value, an
optically thick wind occur (\citealt{HAC96}), where the critical
accretion rate is
 \begin{equation}
 \dot{M}_{\rm c, H}=5.3\times 10^{\rm -7}\frac{(1.7-X)}{X}(M_{\rm
 WD}-0.4) M_{\odot} {\rm yr}^{\rm -1},
  \end{equation}
where $X$ is hydrogen mass fraction and $M_{\rm WD}$ is the mass
of the accreting WD (mass is in $M_{\odot}$ and mass-accretion
rate is in $M_{\odot}{\rm yr}^{\rm -1}$, \citealt{HAC99a}). If
$\dot{M}_{\rm a}$ is smaller than $\dot{M}_{\rm c, H}$, but higher
than $\frac{1}{2}\dot{M}_{\rm c, H}$, the hydrogen-shell burning
is steady, i.e. the system is experiencing a SSS phase. When
$\dot{M}_{\rm a}$ is lower than $\frac{1}{2}\dot{M}_{\rm c, H}$
but higher than $\frac{1}{8}\dot{M}_{\rm c, H}$, a very weak shell
flash is triggered and then a recurrent nova is expected. When
$\dot{M}_{\rm a}$ is lower than $\frac{1}{8}\dot{M}_{\rm c, H}$,
the shell flash is so strong that no material is accumulated on to
the surface of the CO WD\footnote{Actually, the mass of WD is
reduced when a WD undergos a nova explosion.}.

We extract the evolutionary time of Case calm from all the binary
system calculated in \citet{MENGYANG10a} and found that the
typical value for SSS phase is $1.5-4.5\times10^{\rm 5}$ years
which is consistent with the calculation of \citet{HKN10}. Since
\citet{HKN10} have discussed it, we will focus on the relative
duration of SSS phase to the whole mass-increasing time of CO WD
until SNe Ia explosion, $t_{\rm SSS}/t_{\rm SN}$, i.e. the
fraction of the time spent in the SSS phase. Please keep in mind
that as observational X-ray output of a nuclear-burning WD
nonlinearly increases with its mass, i.e. it depends on the
evolutional details of WD binary systems, the results obtained in
this paper are not directly related to the X-ray output of SN Ia
progenitor. However, the results here may partly reveal the X-ray
nature of  SN Ia progenitor at least.

To obtain the distribution of $t_{\rm SSS}/t_{\rm SN}$, we have
performed a series detailed Monte Carlo simulations via Hurley's
rapid binary evolution code (\citealt{HUR00, HUR02}). In the
simulations, if a binary system evolves to a WD + MS or WD + RG
stage, and the system is located in the ($\log P^{\rm i}, M_{\rm
2}^{\rm i}$) plane for SNe Ia at the onset of RLOF, we assume that
a SN Ia is produced. $t_{\rm SSS}/t_{\rm SN}$ is then obtained by
interpolation in the three-dimensional grid ($M_{\rm WD}^{\rm i},
M_{\rm 2}^{\rm i}, \log P^{\rm i}$) of the more than 1600 close WD
binary ysytems calculated in \citet{MENGYANG10a}. In the
simulations, we follow the evolution of $10^{\rm 7}$ sample
binaries. The evolutional channel is described in
\citet{MENGYANG10a}. As in \citet{MENGYANG10a}, we adopted the
following input for the simulations. (1) A single starburst or a
constant star formation rate are assumed. (2) The initial mass
function (IMF) of \citet{MS79} is adopted. (3) The mass-ratio
distribution is taken to be constant. (4) The distribution of
separations is taken to be constant in $\log a$ for wide binaries,
where $a$ is the orbital separation. (5) A circular orbit is
assumed for all binaries. (6)The common envelope (CE) ejection
efficiency $\alpha_{\rm CE}$, which denotes the fraction of the
released orbital energy used to eject the CE, is set to 1.0 or
3.0. (See \citealt{MENGYANG10a} for details).

\section{RESULTS}
\label{sect:3}

   \begin{figure}[h!!!]
   \centering
   \includegraphics[width=9.0cm, angle=270]{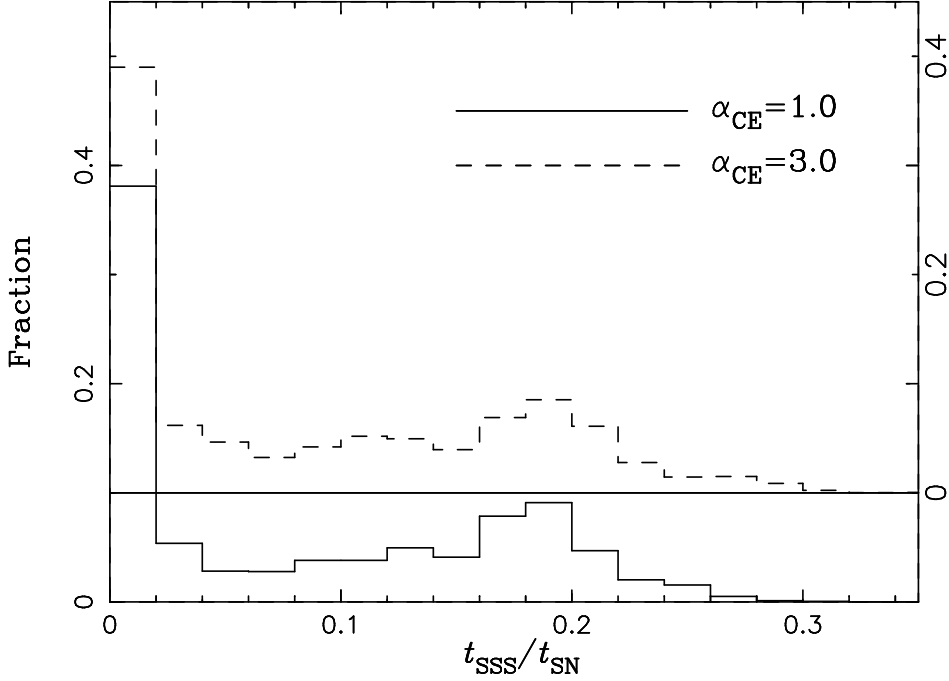}


   \caption{Distribution of the relative duration of SSS phase to
the whole mass-increasing time of CO WDs until SNe Ia explosion,
$t_{\rm SSS}/t_{\rm SN}$, for constant star formation rate with
different $\alpha_{\rm CE}$ (solid line: $\alpha_{\rm CE}=1.0$;
dashed line: $\alpha_{\rm CE}=3.0$). For $\alpha_{\rm CE}=3.0$,
the line is up-moved by 0.1.}
   \label{fretsss}
   \end{figure}

      \begin{figure}[h!!!]
   \centering
   \includegraphics[width=9.0cm, angle=270]{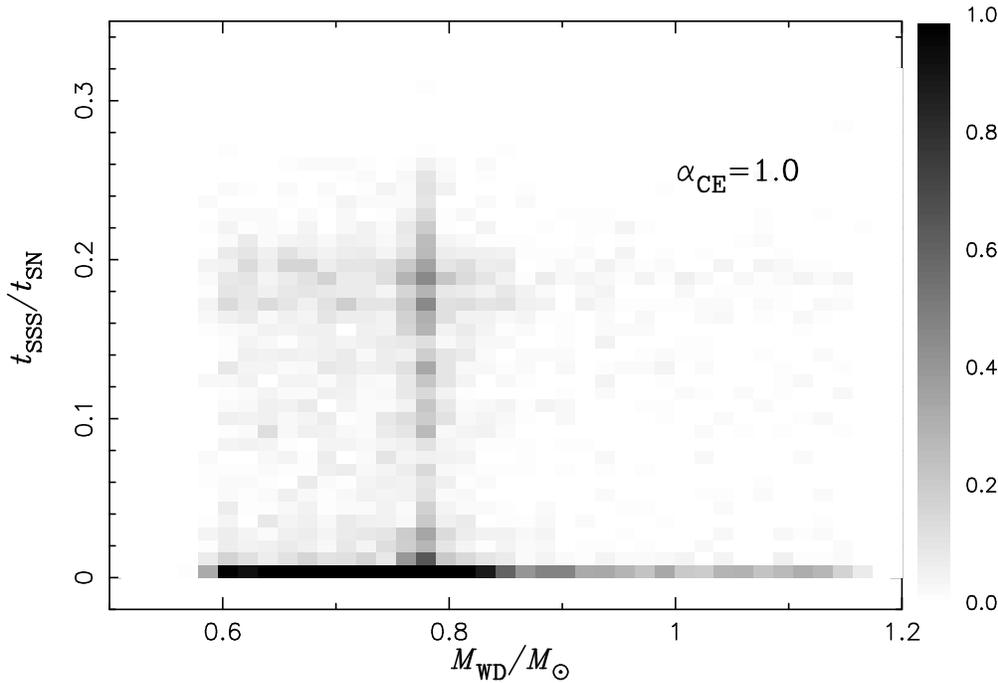}


   \caption{Distribution of the relative duration of SSS phase
and initial WD mass with $\alpha_{\rm CE}=1.0$.}
   \label{ssswd}
   \end{figure}

\subsection{the relative duration of SSS phase}\label{sect:3.1}

In figure \ref{fretsss}, we show the current-epoch-snapshot
distribution of $t_{\rm SSS}/t_{\rm SN}$ for different
$\alpha_{\rm CE}$, where a constant star formation rate is
assumed. We can see from the figure that the distribution is
similar to each other for different the $\alpha_{\rm CE}$. A
remarkable feature in the figure is that it is about 40\% of
mass-accreting CO WDs for SNe Ia that spend less than 2\% of their
mass-increasing time on SSS phase. Actually, for most of these
WDs, the relative duration of SSS phase is only about $10^{\rm
-3}-10^{\rm -4}$. No WD has a relative duration of SSS phase more
than 30\%, which means that if we assume that the WDs show the
signatures of SSSs during their whole mass-increasing time until
SNe Ia, the number of SSSs is at leat overestimated by a factor of
3. In the distributions, there is a smaller bump between 0.16 and
0.2, which correspond to the systems with an initial WD mass of
$0.6\sim0.8 M_{\odot}$, or a primordial primary of
$2.5\sim3.5M_{\odot}$ (see figure \ref{ssswd}, \citealt{MENG08}).
In figure \ref{ssswd}, we give the distributions of the relative
duration of SSS phase and the initial WD mass for $\alpha_{\rm
CE}=1.0$ (The distributions of $\alpha_{\rm CE}=3.0$ is similar to
that of $\alpha_{\rm CE}=1.0$). From the figure, we can see that
whatever the initial WD mass is, the low relative duration of SSS
phase (less than 2\%) is dominant.

   \begin{figure}
   \centering
   \includegraphics[width=90mm,angle=270.0]{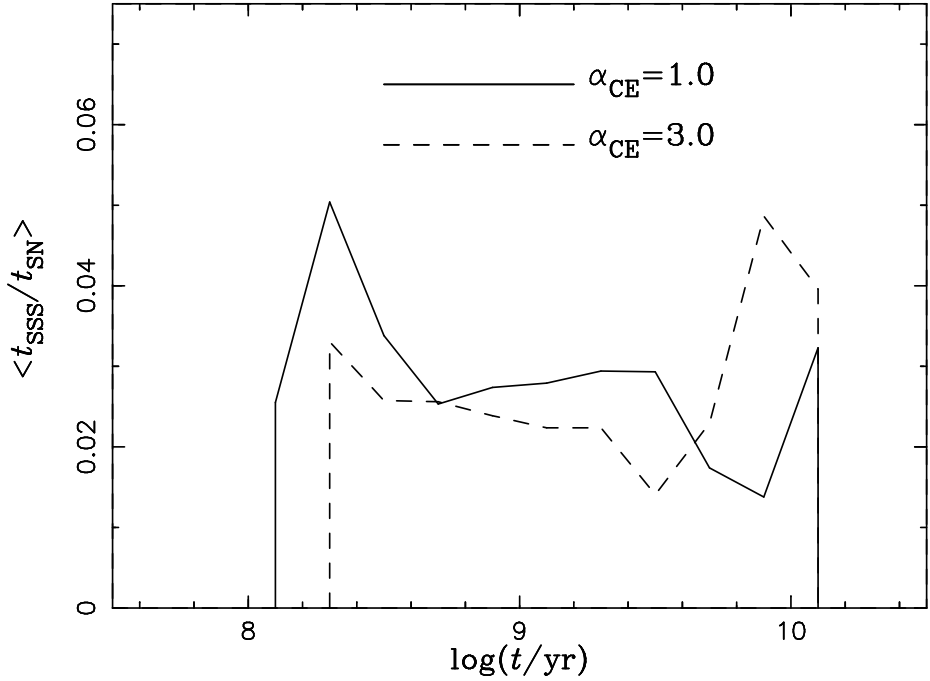}
   \caption{Evolution of the mean relative duration of SSS phase
to the whole mass-increasing time of CO WD until SNe Ia explosion
for a single starburst with different $\alpha_{\rm CE}$ (solid
line: $\alpha_{\rm CE}=1.0$; dashed line: $\alpha_{\rm CE}=3.0$).}
              \label{meanssst}%
    \end{figure}

To explore the contribution of the accreting WD on the soft X-ray
flux of galaxies, a mean relative duration of SSS phase is
necessary. Since the relative duration of SSS phase is covering
two orders of magnitude, we use a geometric mean to denote their
mean value. Figure \ref{meanssst} show the evolution of the mean
relative duration of SSS phase for a single star burst. It is
amazing that the mean relative duration is always smaller than 5\%
from $\sim10^{\rm 8}-10^{\rm 10}$ yrs, which may imply that even
in spiral galaxies, only a small fraction of the mass-accreting
WDs leading to SNe Ia contribute to supersoft X-ray flux.

   \begin{figure}
   \centering
   \includegraphics[width=90mm,angle=270.0]{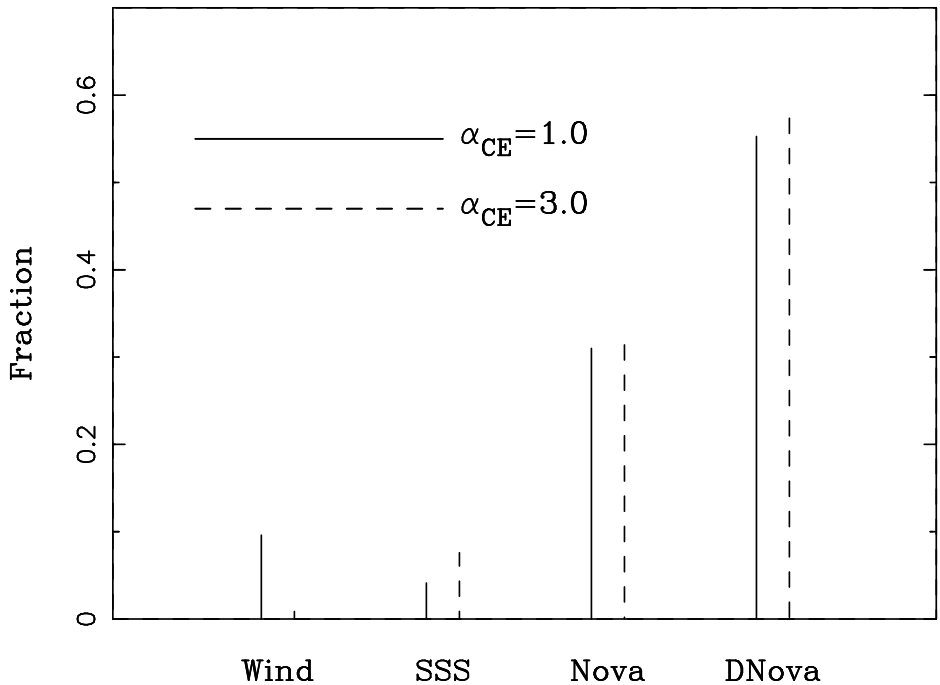}
   \caption{The relative importance of wind, SSS, nova and dwarf nova
at SNe Ia explosion for a constant star formation rate with
different $\alpha_{\rm CE}$ (solid line: $\alpha_{\rm CE}=1.0$;
dashed line: $\alpha_{\rm CE}=3.0$).}
              \label{class}%
    \end{figure}

   \begin{figure}
   \centering
   \includegraphics[width=90mm,angle=270.0]{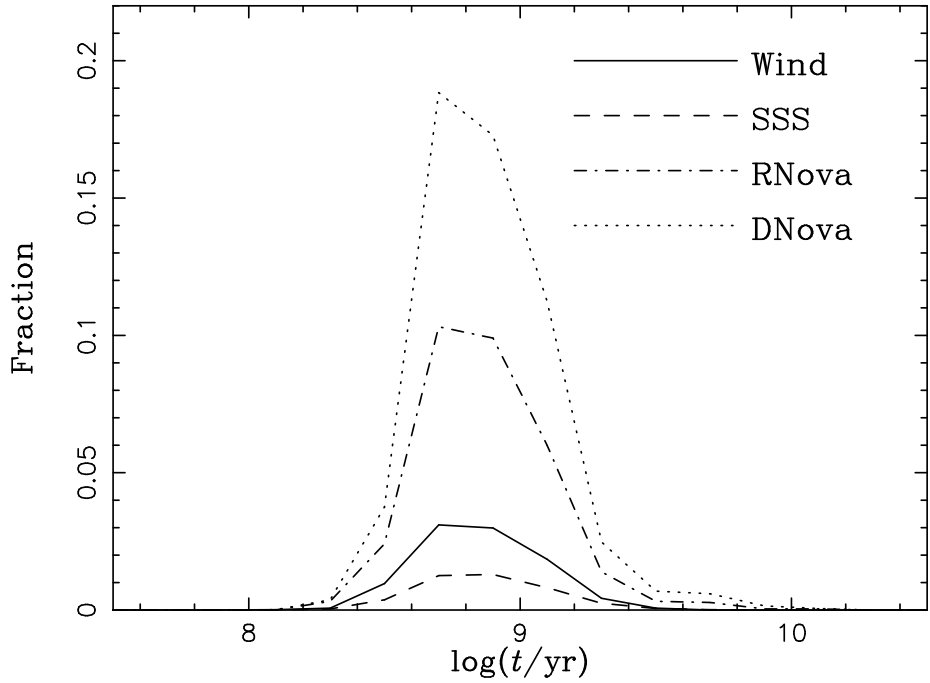}
   \caption{Evolution of the fraction of wind, SSS, recurrent nova and
dwarf nova at SNe Ia explosion for a single starburst with
$\alpha_{\rm CE}=1.0$.}
              \label{ctime}%
    \end{figure}

\subsection{class}\label{sect:3.2}
In the calculations of \citet{MENGYANG10a}, the SNe Ia may explode
at wind, SSS, recurrent nova and dwarf nova phase. In figure
\ref{class}, we show the relative importance of wind, SSS, nova
and dwarf nova at SNe Ia explosion. We can see from the figure
that more than 55\% progenitor systems show the properties of
dwarf nova before SNe Ia, and about 30\% are recurrent novae. Only
no more than 10\% progenitor systems are experiencing the SSS
phase. There are about 2\% -10\% (depending on $\alpha_{\rm CE}$)
SNe Ia exploding at wind phase. These SNe Ia may show the
properties of 2006X-like SNe Ia (see \citealt{MYG10}). Then, even
before SNe Ia explosion, SSS is still not a dominant state.

Figure \ref{ctime} gives the evolution of the fraction of SNe Ia
exploding at wind, SSS, recurrent nova and dwarf nova phase for a
single starburst. Here, we only present the case of $\alpha_{\rm
CE}=1.0$ since the case of $\alpha_{\rm CE}=3.0$ is similar. We
can see from the figure, when age is longer than 2 Gyr, there is
no SNe Ia exploding at SSS phase and almost all SNe Ia exploding
at dwarf nova phase. Considering that supersoft X-ray radiating
from dwarf nova is rare (see discussions in \citealt{HKN10}), the
discovery by GB10, i.e. the supersoft X-ray flux is very low in
early type galaxies, is a natural result.

\section{DISCUSSION AND CONCLUSIONS}\label{sect:4}
In this paper, we give the distribution of the relative duration
of SSS phase and the evolution of the distribution. Our results
show that the duration of the SSS phase for accreting WDs is very
short compared with their mass-increasing time. This result imply
that if we assume that WDs radiate supersoft X-ray during their
whole mass-increasing phase, the number of SSS should be
overestimated significantly. The statement in GB10 that no more
than 5\% of SNe Ia in early type galaxies can be produced by
mass-accreting white dwarfs of the SD scenario is just based on
this simple assumption. Our BPS study shows that the mean relative
duration of SSS phase is always smaller than 5\% for SNe Ia with
both short- and long-delay time, which is consistent with the
discoveries of GB10 and \citet{DISTEFANO10}. Our results mean that
in spiral or elliptical galaxies, the fraction of mass-accreting
WD contributing to supersoft X-ray is very small, as suspected by
\citet{DISTEFANO10}. So, the discoveries of GB10 and
\citet{DISTEFANO10} could be evidence to support the SD scenario
as SNe Ia progenitor model, and then we can not use the low
supersoft X-ray luminosity of galaxies or the low number of
mass-accreting WD showing supersoft X-ray signal in elliptical
galaxies to exclude the validity of SD model.

In addition, please keep in mind that there is another uncertainty
in GB10 that they assume all the supersoft X-ray come from the
mass accreting WDs which will explode as SNe Ia finally. Actually,
the SSS at $t=10^{\rm 10}$ yr are most likely not the progenitor
of SNe Ia, whatever in spiral or elliptical galaxies
(\citealt{YUNGELSON10}). For example, some progenitor systems may
experience a SSS phase, but nova explosion prevents them from
becoming SNe Ia finally (see the figures 3 and 4 in
\citet{MENGYANG10a}). Then, the contribution of mass-accreting WDs
resulting in SNe Ia to supersoft X-ray flux may be smaller than
those shown in this paper, as discussed in \citet{HKN10}.

However, there are some uncertainties for our results in this
paper. Firstly, as well known, a WD radius decreases as its mass
increases. Correspondingly, at the same accretion rate $\dot{M}$
and luminosity, less massive WD would have lower temperature, and
then more soft X-ray (\citealt{VANDENHEU92}). Calculation shows
that WDs less massive than about 1 $M_{\odot}$ do not contribute
to the statistics of SSS (\citealt{DISTEFANO10}). The soft X-ray
emission from galaxies is dominated by the emission from most
massive WDs (maybe close to Chandrasekhar mass limit because of
the special X-ray band, i.e. 0.3-0.7 keV in \citealt{GB10}). For
the reason above, the total luminosity of SN Ia progenitors not
only simply scales with the duration of the super-soft phase but
also depends on the details of their evolutional tracks. Since the
most of SSSs happen when WD masses are around $1.2-1.3 M_{\odot}$
(see \citealt{MENGYANG10a}), it should be taken as an upper limit
of the real cases for the number of SSSs deduced from the results
in this paper. Then, this uncertainty can not change our basic
conclusion, i.e. GB10 overestimates the number of SSSs.

Secondary, a part of accreting ONeMg WDs may also contribute to
the soft X-ray and their contribution was not considered here,
which may make us to underestimate the number of SSSs.
Fortunately, the number of ONeMg WDs must be much smaller than
that of the CO WDs considered here because of initial mass
function. So, the basic conclusions in this paper are still hold.

Thirdly, in this paper, we consider a new possible mechanisms to
SNe Ia, i.e. dwarf nova, which is from a disk instability
(\citealt{OSAKI96}; \citealt{VANP96}). For a low amount of
accreted materials during an outburst, the accreted matter may
simply accumulate onto the WD but remains unburnt. After a few
tens of disk instabilities, the accreted materials finally reaches
a critical mass which may result in a shell flash. This could be
nothing else but a classical (or recurrent) nova (see also
\citealt{HKN10}). This scenario was uphold by observations
(\citealt{WILS10}). However, its SSS duration is too short to be
compared with the steady SSS duration (\citealt{HKN10}). Then, our
result, i.e. the dwarf nova is the dominant progenitors for old
SNe Ia, may imply that the total luminosity of SSSs in early type
galaxies would be much smaller than that predicted by GB10, which
is consistent with the observation of GB10. In addition, I have to
remind the reader that \citet{GB11} recently argued that WD with
unstable nuclear burning, i.e. classical/recurrent nova, may not
be the main class of SN Ia progenitor in early type galaxies
because they would be producing too many recurrent nova outbursts
inconsistent with classical nova statistic in nearby galaxies,
i.e. the observational rate of classical nova with decay time
shorter than 20 days is overestimated by a factor of 45-75. This
inconsistency seems to be another evidence against the validity of
SD scenario in early type galaxies. Actually, in this paper, the
dwarf nova is the dominant one for the progenitor of SNe Ia in
early type galaxies based on the calculation of
\citet{MENGYANG10a}. In \citet{MENGYANG10a}, the duty cycle for
dwarf nova is set to be 0.01 (typical range is from 0.1 to a few
$10^{\rm -3}$), which means that if the dwarf nova is a possible
channel for SNe Ia in early type galaxies, the estimated rate of
classical nova in \citet{GB11} could be overestimated by a factor
of 100 (typical range is from 10 to several $10^{\rm 2}$).
Considering that there are still some recurrent nova contributing
to SNe Ia in early type galaxies, the overestimated factor of
45-75 by \citet{GB11} could be a natural results and should also
not be a key against the SD channel in early type galaxies.

Finally, some mass-accreting WDs may not explodes as SNe Ia and
the contribution of these WDs to soft X-ray luminosity was not
considered in this paper. However, parameter space for these WDs
is much smaller than those exploding as SNe Ia
(\citealt{MENGYANG10a}). Furthermore, the relative duration of SSS
phase for these WDs is much smaller than those exploding as SNe
Ia. So, the missing WDs may not significantly affect our
discussions in this paper.

 \normalem
\begin{acknowledgements}
This work was partly supported by Natural Science Foundation of
China under grant no. 11003003 and the Project of the Fundamental
and Frontier Research of Henan Province (102300410223), the
Project of Science and Technology from the Ministry of Education
(211102) and the China Postdoctoral Science Foundation funded
project 20100480222.
\end{acknowledgements}



\label{lastpage}

\end{document}